# The Imperishable Record, Forgiveness, & the Future of Criminal Justice

**Mark Fenwick** is a professor at Graduate School of Law, Kyushu University, Japan.

**Paul Jurcys** is Senior IP Counsel at Vinted and a senior lecturer at Vilnius University Law Faculty, Lithuania.

**Sebastian Porsdam-Mann** is Assistant Professor at the Center for Advanced Studies in Bioscience Innovation Law (CeBIL) at the University of Copenhagen and a Visiting Associate Professor at the National University of Singapore's Centre for Biomedical Ethics.

## Abstract

Digital technologies can preserve records of wrongdoing indefinitely, and AI can give those records continuing influence over a person's future. We describe this condition as the eternal data present and argue that it threatens the possibility of forgiveness. In criminal justice, formal punishment has an end, whereas the harms caused by the continued use of criminal records may not. Drawing on Hegel's account of the "hard heart" and Arendt's account of the irreversibility of action, we distinguish two dangers that digital systems can combine. People may be permanently identified with what they have done, and acts that cannot be undone may continue to determine what happens to them. We propose a right to remain unfinished, supported by rights to limits on the weight given to older records, to re-evaluation in light of personal change, to reassessment of inferences that affect their lives, and to time limits on the use of records in specified decisions. These proposals extend the ethics of AI beyond accuracy, transparency, and non-discrimination to address how long past wrongdoing should be allowed to govern a person's life.

# Table of Contents

# 1. Introduction

How long should a past wrong continue to define a person? This is one of the oldest ethical questions confronting any formal system of punishment. Modern criminal justice necessarily remembers, recording the wrongful act, identifying the person responsible, and attaching consequences to what has been done. Without such institutional memory, responsibility itself would be difficult to sustain. Yet criminal justice also assumes that punishment should, at some point, end. Sentences expire and prisoners are released; convictions may become spent, and records may be expunged or sealed. Retribution, rehabilitation, and reintegration presuppose, in different ways, that although the past cannot be changed, its effects on the present and future need not remain constant.

Digital technologies make these limits harder to maintain by increasing the ability of criminal justice institutions, private organisations, and ordinary citizens to preserve, retrieve, combine, and act upon information about the past. Arrests, convictions, allegations, disciplinary incidents, police intelligence, risk assessments, and other records of criminal justice involvement may remain accessible long after any associated formal punishment has ended. Remembering has become easier and forgetting increasingly exceptional (Mayer-Schönberger 2009).

AI may intensify this problem by using historical information to classify people, identify correlations, and draw inferences about them. Past actions become inputs into predictions about future dangerousness, recidivism, reliability, compliance, or trustworthiness. A record of what happened can thereby help determine how a person is understood, even after the formal consequences of the deed have ended.

Work on digital punishment and collateral consequences has shown how criminal records can continue to affect access to employment, housing, and other opportunities long after a formal sentence has ended (Lageson 2020). These continuing consequences are also the concern of scholarship on legal forgiveness, which asks when law should release individuals from them through practices such as pardon, amnesty, expungement, restorative justice, and other forms of institutionalised second chances (Minow 2019). Limits on the information itself provide another response. Privacy scholarship has focused on the right to be forgotten, erasure, and other mechanisms for limiting the persistence or accessibility of personal information (Solove 2008, 2025a). Where records remain in use, debates over AI in criminal justice have concentrated on predictive accuracy, discrimination, transparency, explainability, and procedural fairness (Huq 2019, Mayson 2019).

These responses leave a further question. How long should historical information influence judgments about a person who may have changed, even when that information is accurate, legitimately retained, and fairly processed? We describe this continuing influence of the past as the *eternal data present*. We argue that digital technology threatens forgiveness by preserving records of criminal acts indefinitely, while AI enables those records to continue influencing an individual's future.

This question concerns forgiveness because it offers a way of acknowledging the past without allowing it to govern the present or future indefinitely. Although it neither denies the deed nor dissolves responsibility, it rejects the conclusion that a person should be understood entirely in terms of that deed.

Hegel's discussion of the "hard heart" and forgiveness toward the end of the *Phenomenology of Spirit* concerns the relationship between the person who acts wrongly and the person who judges (Hegel 1977, §666-671). The hard-hearted judge treats the deed as though it disclosed the complete truth of its author. Forgiveness breaks this identification by allowing the deed to remain attributable to the subject without exhausting who that subject is.

Hannah Arendt's discussion of forgiveness in *The Human Condition* addresses the consequences of a deed that cannot be undone (Arendt 1958, pp. 236-243). She approaches the problem through the irreversibility of action, since once something has been done, it cannot be made not to have happened. Without forgiveness, our capacity to act would itself be, in Arendt's phrase, "confined to one single deed" from which recovery would be impossible. Forgiveness allows a past that cannot be undone to cease determining everything that follows.

These perspectives identify related but distinct dangers. The Hegelian danger is expressed in the judgment "This is what you did, therefore this is what you are." The Arendtian danger is expressed in "This is what you did, therefore this is what will continue to happen to you." Digital criminal justice risks combining them. Because a record remains available, a deed may come to define how the person is understood, while algorithmic systems allow that assessment to determine future decisions about them.

We propose a right to remain unfinished, which would prevent any criminal record, profile, or computational representation from acquiring final authority over who a person is or what they may become. After examining the persistence and use of criminal records in Section 2, we develop the account of forgiveness in Section 3 and apply it to predictive systems in Section 4. Section 5 gives the proposal institutional form through rights to limits on the weight of older records, to re-evaluation, to reassessment of inferences, and to time limits on specified uses. A criminal justice system may retain the records it needs while recognising that the people described in those records can change.

## 2. The Eternal Data Present & Modern Criminal Justice

### *2.1. The Persistence of Digital Records*

Memory is, and always has been, constrained by material and social conditions (Stiegler 2009). Some forgetting was therefore inevitable in ordinary social life, although it did not amount to systematic or morally principled forgiveness. Forgetting could conceal serious wrongdoing just as easily as it could allow a person to begin again.

Digital technologies make forgetting less likely. Information can now be preserved at almost negligible marginal cost, reproduced without degradation, searched at speed, and distributed across different contexts (Mayer-Schönberger 2009; Ohm 2010). Under these conditions, the past remains continuously available for use in the present (Fertik, Fenwick and Jurcys forthcoming). The distinction between "then" and "now" can become difficult to sustain when records are used without regard to the time that has elapsed.

Continuous availability matters when information created at one point in a person's life serves as evidence about them at another. People may change while their records remain the same. A record can therefore accurately represent someone as they once were while giving a deeply misleading impression of who they are now.

For example, when a record's age no longer qualifies the judgment drawn from it, "This person said X in 2008" becomes "this person believes X today." Although the historical statement remains accurate, the judgment about the present may be misleading. A record may remain true without continuing to justify the conclusions once reasonably drawn from it.

One response is to limit the availability of the information itself. Debates about digital privacy have often focused on forgetting through deletion, erasure, de-indexing, record sealing, and related mechanisms (Mayer-Schönberger 2009; Solove 2006). These responses limit how long information remains available, alongside proposals to give individuals meaningful control over the data generated by their own lives (Jurcys et al. 2020; Donewald et al. 2021; Fenwick and Jurcys 2022; Fenwick et al. 2025). Where information remains available, a different question concerns the judgments that may be drawn from a remembered deed. Forgiveness addresses those judgments, which can change without the information being removed.

The judgments drawn from a record can also be sustained by its repeated circulation. Online platforms create strong incentives to bring past actions back to public attention, exposing them to repeated judgment long after their original social context has disappeared. The behavioural literature on availability cascades explains how a perception gains plausibility as it becomes more available in public discourse, and how individuals endorse it partly by learning from the apparent beliefs of others and partly to protect their own standing (Kuran and Sunstein 1999). People may come to remember the most widely circulated account of a deed, which becomes easier to retrieve each time it is repeated.

Repeated exposure can thus keep an old deed before an audience. Business models that depend on attracting attention may also reward content that provokes anger, surprise, moral indignation, or conflict (Zuboff 2015). Public exposure of wrongdoing can be necessary, particularly where powerful actors have escaped accountability. Appeals to forgiveness should not be used to instruct victims to remain silent or tolerate repeated abuse.

The difficulty arises when this continuing exposure fixes how a person is understood by others (Solove 2008; Fertik, Fenwick & Jurcys). A recorded action can support inferences that come to define the person who performed it. Although drawing conclusions about people from their conduct is often appropriate, those conclusions should not become permanent or exhaustive.

Criminal justice presents this difficulty in an institutional setting explicitly committed to remembering. It produces records precisely because actions matter over time. Arrest histories, convictions, sentencing remarks, prison records, probation reports, police intelligence, risk assessments, and other documentation preserve information about conduct that the state considers significant.

Preserving these records serves legitimate purposes. Prior convictions may be relevant to sentencing, and police may need historical information to investigate patterns of offending. Courts need records of previous proceedings, while correctional institutions require information about the people in their custody. Victims and the public may also have legitimate interests in preserving serious criminal histories.

The legitimacy of keeping these records, however, does not in itself imply that their consequences should be unlimited. Criminal justice has traditionally limited how long the past may continue to affect a person. Sentences end, and parole terminates. Convictions may become spent, juvenile records may receive special protection, and expungement and sealing regimes restrict future access. Rehabilitation remains an institutional objective, although commitment to it is declining in many jurisdictions (Garland, 2001). Even retributive punishment commonly assumes a point at which the person has discharged the debt imposed by the sentence.

### *2.2. Digitalisation*

Digitisation makes it harder to maintain these limits, especially where the consequences of a criminal record extend beyond the formal sanction. A criminal sentence specifies a period of imprisonment or supervision, a financial penalty, or another legal consequence. The harms caused by the continued use of a record are harder to identify and delimit. A sentence may end while the record continues to affect employment, education, housing, credit, migration, insurance, personal relationships, and interactions with state institutions (Lageson 2020).

These effects beyond the formal sentence are often described as collateral consequences (Kirk and Wakefield, 2018). Some are legally prescribed; others arise through private decisions or informal judgments. In each case, information produced within criminal justice continues to be used after the formal legal process has ended.

The completion of a legal sentence therefore need not bring practical consequences to an end. Criminal records can be retrieved rapidly, linked across databases, and incorporated into other information systems (Lageson 2020). Records that might once have remained in an archive are therefore more readily put to use, making the distinction between retaining a record and acting on it harder to maintain.

The traditional account of punishment envisages a sequence ending in reintegration.

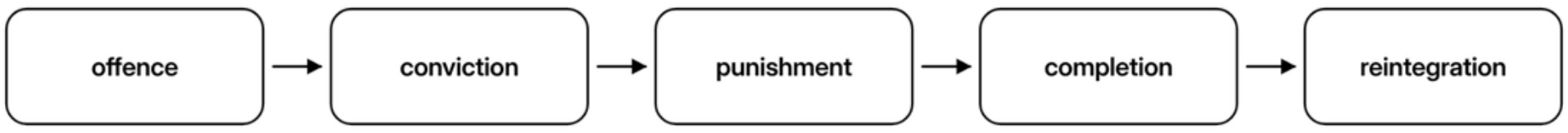

Repeated use of digital records can instead produce a recurring sequence.

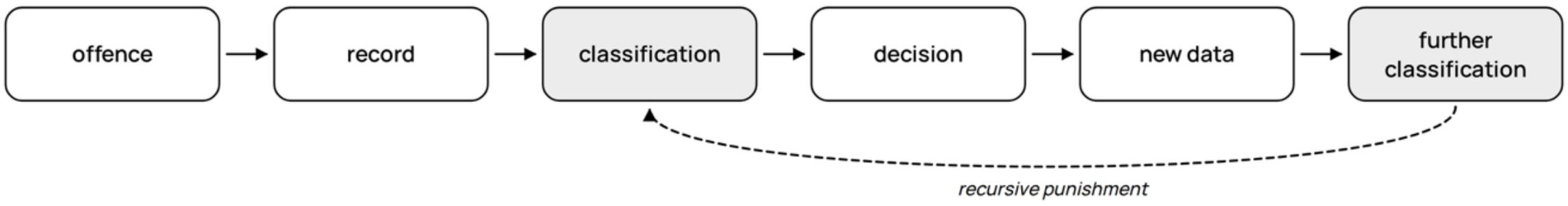

The second sequence has no obvious end point, so the consequences of criminal justice involvement can recur. An initial classification affects later opportunities, which may in turn affect a person's contact with institutions. Each interaction generates further data. A person once designated as risky may therefore repeatedly encounter systems that treat the original designation as evidence of continuing risk.

This repetition can extend beyond the institution that made the original classification, since a classification made for one purpose can be used for another. A judgment about risk in criminal justice may become a broader judgment that a person is unreliable, influencing decisions about employment, housing, or credit even where the original offence has little relevance to the decision.

When these classifications continue to affect people in settings formally outside punishment, we must ask when keeping and using a legitimate record of wrongdoing becomes a means of punishing someone indefinitely.

The answer cannot turn on the existence of a disadvantage alone, since not every disadvantage following conviction amounts to punishment and other institutions may have legitimate reasons to take certain histories into account. Forgiveness need not require the immediate restoration of every status after every offence. Some conduct may remain relevant for a long period, particularly where a future decision involves a closely related risk.

## *2.3. Institutional Records & the Uncertain Ethics of Punishment*

One question that arises from these considerations is how institutions should decide whether continued reliance on a record remains justified. A criminal justice system can formally endorse rehabilitation while continuing to use records in ways that undermine it. A person may complete every legally required element of punishment yet repeatedly have an old deed presented as evidence of their present character. Whether this is justified depends on how much time has passed, the relevance of the deed, and whether the person has changed.

Rehabilitation requires both personal change and recognition of that change by others. A person may change and become less likely to offend, and still have to persuade institutions to revise their assessment. The individual can therefore change while the institutions using their records continue to treat them in the same way (McNeil 2012).

Forgiveness addresses the institutional side of this issue. Rehabilitation usually concerns whether the person who committed the wrong has changed; forgiveness also concerns whether those judging them are capable of revising their judgment. Institutions must at some point reconsider how they treat the person in light of the deed, so the burden of change should not lie entirely with the person who committed the wrong.

A retributive approach might appear to avoid this dependence on forgiveness by holding that wrongdoing deserves a specified amount of punishment. Once that punishment has been endured, the debt has been paid, whether or not the person has been forgiven. Forgiveness may therefore seem conceptually unnecessary. Yet if institutions continue indefinitely to treat the person as morally tainted after the formal debt has been discharged, the completed punishment has become a basis for condemning the person permanently.

Behavioural evidence suggests why such judgments may persist. Punishment judgments track the outrageousness of the act rather than a calculation of what punishment will achieve. Experimental subjects, including judges, hold to that intuition even when told that the probability of detection was low or that the sanction would leave people less safe (Sunstein 2005). An outrage heuristic of this kind offers no basis for deciding when condemnation should end. Although retributivism limits the severity of punishment, it does not by itself limit how long digital records continue to affect a person.

For rehabilitation, the difficulty concerns how change is recognised. If forgiveness is deferred until the person demonstrates change, that evidence must be able to affect subsequent assessments. When every new decision relies on a system that repeatedly returns to the original wrongdoing, evidence of change may never receive sufficient weight. Rehabilitation can then remain possible in principle while being difficult for institutions to recognise.

Both approaches must explain how punishment can end when records continue to be used. Release from custody need not end a person's classification as an offender, and completing a sentence need not end the consequences of the record it generates. A digital criminal justice system therefore needs limits on how long those records may continue to determine the person's treatment.

The question is how to impose these limits while retaining records that serve legitimate purposes. Such limits need not require the deletion of criminal justice data. A conviction can remain in the historical record while carrying less weight in assessments of the person. Without a way to reduce that weight, a system may demand that an offender change while making it progressively harder for evidence of change to affect how they are treated.

# 3. Forgiveness, Punishment and the Possibility of a New Beginning

## *3.1. Hegel's Hard Heart*

Forgiveness can limit the authority of a remembered deed without cancelling responsibility for it. Hegel and Arendt offer complementary accounts of how this is possible, distinguishing responsibility for a deed from permanent subjection to it.

Hegel's discussion appears near the end of the *Phenomenology of Spirit*, in the sequence concerning conscience, the beautiful soul, evil, and forgiveness (Hegel 1977, §666-671). The scene involves two consciousnesses: one that has acted and comes to acknowledge its wrongdoing, and another that judges that wrongdoing. Through confession, the person who acted says, in effect, "Yes, I did this." In acknowledging the deed as their own, the wrongdoer accepts responsibility and opens themselves to the other.

Acknowledging wrongdoing does not by itself reconcile the two parties, since the confession also appeals for reciprocity. By admitting the deed, the actor implies that the two consciousnesses are the same in kind and invites the judge to acknowledge as much in return. The judging consciousness refuses, maintaining a moral distance from the actor. Hegel calls this the *hard heart*, which withholds the answering acknowledgement and insists on a rigid division between the compromised actor and the pure observer. Hegel treats this division as spurious and the judge as a hypocrite because judging is itself a deed, issued from a particular standpoint. The judge therefore shares the partiality it condemns.

Although the actor initially appears to occupy the ethically defective position, confession changes the relationship. The person who acted has acknowledged the deed and opened themselves to reconciliation, whereas the judge obstructs reconciliation by refusing to recognise that the actor may stand in a different relationship to what they have done. The judge treats the person as defined by their wrongdoing.

The judge's refusal becomes objectionable when it treats a deed as the whole truth of a person, rather than attributing it to them in the first place. Hegel does not suggest that responsibility can simply be dissolved. A deed expresses something about its author, but people can subsequently reflect on their actions, reinterpret them, reject previous reasons, confess, change commitments, and act differently.

Hegel's figure of the "beautiful soul" explains why the judge's claim to purity is unstable. The hard heart is the beautiful soul in the posture of judgment. The beautiful soul tries to preserve its moral purity by declining to act, since any action would commit it to something particular and expose it to judgment in turn. It therefore judges without entering the world in which judgment can be answered.

Forgiveness occurs when the judge relinquishes this claim to purity and the hard heart breaks. Hegel identifies this with the movement the confessing consciousness has already performed, as the judge acknowledges its own particularity rather than granting a concession from above.

Hegel writes that “the wounds of the Spirit heal, and leave no scars behind” and that “the deed is not imperishable” (Hegel 1977, §669). The event does not thereby vanish from history, but the deed ceases to provide the final account of who the subject is.

Forgiveness thus requires both parties to change. Through confession, the wrongdoer relinquishes a purely private relationship to the act. The judge gives up both the finality of the verdict and the claim to purity from which it was issued, making a kind of confession in its own right. This reciprocal recognition makes reconciliation possible because the person remains responsible *for* the deed without being defined entirely *by* it.

### *3.2. Arendt on Forgiveness*

While Hegel explains how the relationship between actor and judge can change, Arendt addresses the continuing effects of the deed. In *The Human Condition*, she presents forgiveness as a response to one of two predicaments intrinsic to action. Once an action has entered the world, it cannot be recalled or undone. It is also unpredictable, since the actor cannot control the processes their initiative sets in motion. Arendt pairs each predicament with a faculty drawn from action itself, with forgiving responding to irreversibility and promising to unpredictability. Our argument concerns the first. The problem is not principally epistemic, since an action has become part of the world and continues to have effects beyond the actor’s control, regardless of whether it is remembered.

For Arendt, forgiveness provides a remedy within human action itself. Without it, actors would remain indefinitely bound to what they had once done, making action intolerable. Our capacity to act would be “confined to one single deed” from which recovery would be impossible (Arendt 1958, p. 237).

The contrast with vengeance explains how forgiveness can interrupt these consequences. Vengeance remains determined by the wrong. One deed generates another in response, which provokes a further reaction, allowing the original action to continue shaping and justifying what follows. Forgiveness does not make the first action disappear, but it introduces a response that is not dictated by that action.

Arendt connects forgiveness with freedom understood as the capacity for *new beginnings*. A person can remain connected to the past while initiating something that it does not completely determine. Her formulation that “what was done is forgiven for the sake of who did it” (Arendt 1958, p. 241) distinguishes the *what* of the deed from the *who* of the person. This distinction preserves responsibility while allowing the person to be understood as more than the action.

The capacity to end the consequences of a wrong also explains why Arendt treats punishing and forgiving as alternatives rather than opposites. She also recognises limits to each, holding that some offences admit of neither punishment nor forgiveness and therefore fall outside the reach of human affairs altogether (Arendt 1958, p. 241). Her concern with ending the consequences of action complements Hegel’s concern with how one person recognises another.

Hegel's concern is that a deed should not finally determine who a person is, while Arendt's is that it should not indefinitely determine what happens to them. Both concerns arise in criminal justice, where law attributes a deed to a person and imposes consequences over time. Even when the attribution and initial consequences are justified, the danger remains that the deed will permanently define the person and govern their future.

Applying these accounts to criminal justice requires a further step, since forgiveness is sometimes regarded as too private, emotional, or theological to inform public institutions. Criminal law is typically organised around responsibility, culpability, punishment, deterrence, rehabilitation, and public protection, whereas forgiveness may appear to concern interpersonal relations. Yet legal systems already contain practices through which justified grievances are relinquished. Martha Minow defines forgiveness in these terms and identifies pardon, commutation, amnesty, bankruptcy, and the discretion of police, prosecutors, and judges as tools through which legal systems already forgive (Minow 2019).

Rehabilitation presumes that an offender can change, while reintegration presumes that the community can revise its relationship to someone who has been punished. Spent-conviction rules limit how long past wrongdoing remains relevant, and expungement and record sealing create legal conditions for a fresh start. Restoring civic rights marks a transition from being treated as an offender to fuller membership of the community.

These practices have different legal and political motivations, and none is identical to forgiveness. They nevertheless share a refusal to let criminal wrongdoing justify disadvantages without limit. Forgiveness also differs from exoneration. Where a person was wrongly convicted, the appropriate response is to correct the record and vindicate them. Forgiving someone presupposes that they did something wrong. Nor is forgiveness an excuse, which weakens or removes responsibility for a deed; forgiveness applies where responsibility remains.

Forgiveness need not involve forgetting a deed, which can remain accurately recorded and morally serious without justifying every subsequent judgment about the person. Once responsibility is acknowledged, punishment may itself help limit further condemnation by providing a socially recognised response to the offence. A completed sentence can therefore make continued condemnation less justified, creating conditions under which punishment and forgiveness are compatible.

Digital systems complicate this possibility because a legal sentence may end while judgments based on the record continue indefinitely. Criminal justice may therefore need institutions through which those judgments can be revised. The state need not "feel" forgiveness to make such revision possible. In Hegel's terms, those judging must be able to cease identifying the person with the deed; in Arendt's, the person must be able to cease being governed by its consequences. A criminal justice system capable of forgiveness must satisfy both requirements.

## 4. AI, Prediction & the "Technological Hard Heart"

Predictive systems test whether institutions can meet both requirements, because they connect judgments about the person to decisions about their future. Digital records expand what can be remembered (Ernst 2013), while AI allows that information to be used to make predictions. Whereas a conventional archive stores information about what happened, a predictive system uses past conduct to generate claims about what a person is likely to do next.

Inferring future conduct from the past is not specific to AI. Human decision-makers have long inferred character and risk from previous behaviour. Courts consider criminal histories, police form suspicions, parole boards assess the likelihood of reoffending, and employers judge trustworthiness.

AI changes the speed and scale of these judgments while lending them an appearance of systematic assessment. By processing many data points simultaneously and identifying relationships among variables, AI systems can produce scores and classifications through which institutions assess a person's past.

The Hegelian concern arises when a statistical relationship becomes a description of the person, expressing the judgment "You did this, therefore this is what you are." A series of prior incidents may be treated as evidence that someone is inherently dangerous, high-risk, antisocial, unreliable, or non-compliant, even though the system has identified a pattern in behaviour.

A model need not explicitly condemn someone or make any claim about their moral character for its classifications to function socially as character judgments. In practice, the distinction between "this person has a calculated probability of reoffending above threshold X" and "this is a dangerous person" may be lost. Once a probability informs decisions about a person, it can become a reason for treating them as someone who warrants greater control.

Once the classification guides treatment, the Arendtian concern also arises, expressed in "You did this, therefore this is what will continue to happen to you." Predictions derived from a past deed can affect a succession of decisions about bail, sentencing, parole, supervision, policing, or access to opportunities. When decisions based on a classification generate further records, the two dangers can reinforce each other. Historical criminal justice involvement contributes to a risk classification (Garland 2001), which may lead to greater surveillance or more restrictive treatment. The resulting scrutiny generates additional data that reinforce the original classification (Lageson 2020).

We use the metaphor of the *technological hard heart* to describe a system that continues to judge someone through their history. Algorithms do not possess hearts, moral attitudes, or the human capacity to forgive; the concern is how judgments persist over time.

Reconsideration is possible, but systems must allow for it. Human judgments can change as people forget, shift their attention, form new relationships, or encounter experiences that alter their previous evaluations. A decision-maker who once regarded a person with suspicion may come to recognise evidence of change. Computational systems revise their assessments in these ways only when designed to do so. Otherwise, a classification can remain in use after the reasons for applying it have weakened, simply because the system continues to treat the historical data as relevant.

A technological hard heart reproduces judgments about the past without a mechanism for reconsidering whether they remain justified. Criminal justice must often make judgments under uncertainty, and some historical information can legitimately inform them. A recent pattern of serious offending may matter to a decision about risk, so ignoring all past behaviour would be neither rational nor responsible.

Predictive AI models generally derive their authority from patterns in historical data, whereas forgiveness and rehabilitation depend on recognising that a person's future may depart from those patterns. Assessments must remain open to changes in a person that their previous conduct would not have led us to expect.

A justice system organised entirely around prediction would risk treating such change as noise. The concern becomes more serious when predictions affect the conditions that make change possible. Employment, housing, and social stability are often associated with successful reintegration. If a prediction of risk derived from a criminal record helps exclude someone from these opportunities, the system may contribute to the conditions it claims to predict.

This does not mean that every prediction is self-fulfilling. It means that prediction must be considered within the institutions that help determine how a person's life develops. Accuracy alone cannot resolve the resulting ethical concern. A model might correctly identify a higher average probability of reoffending among people with certain historical characteristics while giving insufficient weight to evidence that a particular person has changed.

The difficulty persists alongside the problems addressed by familiar approaches to AI ethics. Requirements for transparency concern whether we know how a decision was made, while accuracy concerns how well a prediction performs. Non-discrimination addresses unfair distributions of burdens across protected groups, and procedural fairness concerns whether affected individuals can understand or contest the process (Solove 2025b).

These requirements are essential, but none directly determines how long the past should remain a basis for predicting a person's conduct. One objection to the technological hard heart metaphor is that algorithmic decisions are in some respects more open to correction than the human judgments they replace. A model's inputs, objective function, and trade-offs can be inspected in ways that a parole board's reasoning cannot, making algorithms "not only a threat to be regulated" but also a possible instrument of equity (Kleinberg et al. 2018). We accept this premise. Precisely because a system's treatment of the past is specified, its duration can also be specified. A human judge cannot be made to stop weighing what she remembers, whereas a model can be built so that a classification expires unless it is deliberately regenerated.

The technological hard heart results from choices about system design, so regulation can require those choices to change. The same need for reconsideration extends beyond individual risk classifications to the prospect of AI systems constructing detailed profiles from long histories of behaviour and interaction (Jurcys et al. 2026). Such profiles may eventually amount to dynamic models claiming to represent an individual's tendencies in making decisions, preferences, or dispositions.

Such a model represents a person through evidence of their past behaviour. Even if accurate when constructed, it may preserve continuities that no longer describe someone who has changed. It can therefore become more accurate as an account of *who someone was* while becoming less reliable as an account of *who they are*. Criminal justice institutions should be particularly cautious about treating the model as a substitute for the person. The state's legitimate interest in remembering wrongdoing does not give it unlimited authority to define a person by that wrongdoing.

## 5. From Forgiveness to New Rights

If a system can be designed to reconsider an earlier judgment, the remaining question is when institutions should be required to do so. Giving forgiveness a role in digital criminal justice requires institutional principles, since in the interest of justice and consistency criminal justice cannot rely solely on discretion or mercy. Where bias is predictable, law does better to establish institutions that counteract it than to ask individual decision-makers to overcome it (Jolls and Sunstein 2006). We propose a *right to remain unfinished*, supported by rights and duties that preserve the distinction between a person and their record, and between a past deed and what they may do in the future.

This right would prevent any criminal record, profile, or computational representation from acquiring final authority over who a person is or what they may become. It would neither deny history nor remove responsibility for wrongdoing. Four more specific rights would give effect to this principle.

*Image 1. New Rights*

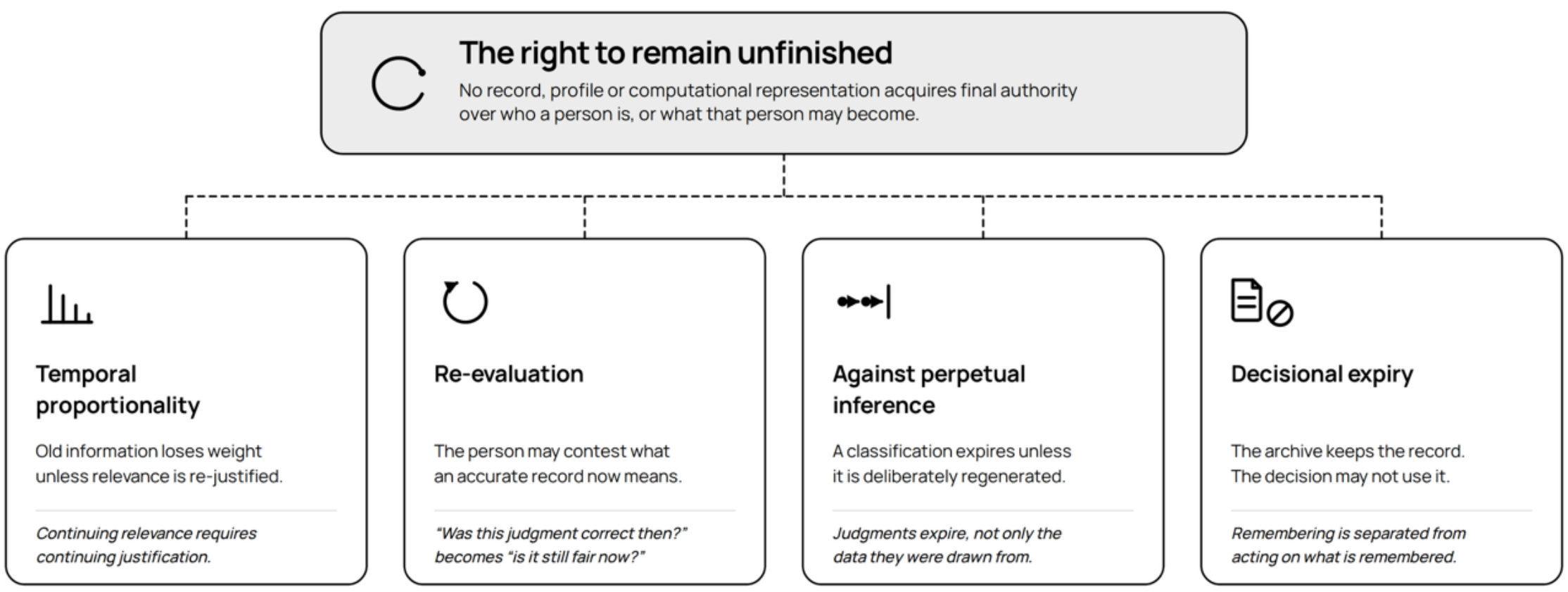

## *5.1. A Right to Limits on the Weight of Older Records*

The first right addresses the difference between a record remaining true and remaining equally relevant. Criminal justice already requires punishment to be proportionate. The weight given to criminal records should also reflect the time that has passed, since a conviction may be highly relevant immediately after an offence but become progressively less informative after years of lawful conduct.

The weight that institutions may give to criminal justice information should ordinarily diminish over time. No universal formula would determine how quickly this should occur, since offences and contexts differ. Information used to protect vulnerable persons may justifiably remain relevant for longer than information used for an unrelated purpose.

If the relevance of an old event can diminish, institutions should have to justify their continued reliance on it when making decisions with present consequences. The availability of a record can reverse that burden, allowing it to remain in use until someone shows why it should not. Defaults encourage inertia and establish a reference point from which departure appears costly (Sunstein 2013). Requiring justification for continued use would place the burden on the institution to explain why the old event remains sufficiently relevant.

Diminishing the weight of an old event must also leave room for subsequent conduct to affect the assessment. A person who has lived for 15 years without further offending and has contributed positively to society should not be assessed as though those years count for nothing. Information about the past may remain true without deserving the same weight in every later decision.

## *5.2. A Right to Re-Evaluation*

Reducing the weight of older records does not by itself provide an opportunity to show how a particular person has changed. The second right would allow a person to seek re-evaluation of an assessment even when the information on which it rests is accurate. Existing data rights often allow challenges to information that is false, incomplete, or inaccurately processed (Solove 2008; Solove 2025b). Forgiveness is relevant where none of those defects is present.

A conviction may have occurred and a risk assessment may have been justified at the time, while the same assessment is no longer justified now. A person should therefore be able to accept the historical record while contesting its continuing significance. Re-evaluation would address both “Was this judgment correct then?” and “Is it still fair to make this judgment now?”

Re-evaluation would allow a person to demonstrate change through later conduct, rehabilitation, treatment, employment, education, and other relevant evidence. It would guarantee that the assessment is reconsidered, without guaranteeing a favourable outcome. In Hegelian terms, those who judge must be willing to revise their relationship to the person they have judged.

### 5.3. A Right to Have Inferences Reassessed

An opportunity for a person to seek re-evaluation also leaves open how institutions should handle inferences that remain in circulation. The third right would prevent inferences drawn from past conduct from remaining in use indefinitely without reassessment. A conviction records an event, whereas a conclusion that the person remains dangerous is a judgment drawn from that event. In algorithmic systems, such a judgment may persist within a profile, feed later models, or influence subsequent assessments until it lasts as long as the underlying record.

Inferences that affect a person's life should remain in use only for a specified period, and risk classifications should be periodically regenerated with attention to subsequent conduct. Models should distinguish historical facts from current assessments, so that an old designation of "high risk" does not become a permanent attribute.

Periodic reassessment must also reach classifications used as inputs to later judgments. When one risk classification contributes to another, the earlier score should not automatically support the later one without scrutiny of its basis and whether circumstances have changed. The survival of the underlying data does not justify preserving a judgment indefinitely. Forgiveness is relevant to AI design because systems can be built to recognise that earlier judgments are no longer justified, without requiring machines themselves to forgive.

### 5.4. A Right to Time Limits on the Use of Records

Reassessing an inference leaves a further question about whether a record should continue to be used in a particular class of decisions at all. The fourth right would place time limits on the use of records in specified decisions. Criminal justice institutions may legitimately retain records for accountability, victims' interests, and research, without thereby being entitled to use every record in every future decision. Keeping information in an archive and using it to decide how someone should be treated require separate justification.

Information could remain archived while its use in specified consequential decisions is prohibited or heavily restricted. The state would retain a record of the event without an unlimited entitlement to act on it. Forgiveness could therefore require an end to a particular use of information while allowing the record to remain intact, preserving history without letting it determine that future decision.

This separation of retention from permitted use builds on existing legal limits. Spent-conviction and expungement regimes already reflect aspects of this principle (Minow 2019). Extending it to AI would require decisions about both the retention of data and their continued use in judgments about people. Because AI makes reuse cheap and possible on a large scale, legal limits may be needed to restore some of the protection once provided by the practical difficulty of retrieving paper records.

Each of these rights would require institutions to justify a different aspect of their continued reliance on the past. An old record could carry less weight; a person could seek re-evaluation; an inference would require reassessment; and a specified use could cease even while the underlying information remained archived. The continued existence of a record would no longer suffice to justify its continued influence over a person's life.

## 6. Future Directions

Criminal justice institutions should be able to retain accurate records while revising the judgments drawn from them. A person may still have a future in chronological terms while losing the possibility of having that future understood independently of their past. The rights proposed here would give legal and technical effect to a broader right to remain unfinished, preventing historical information from becoming a final account of personal identity.

Accuracy, transparency, and non-discrimination remain necessary, but an accurate statement about a person's past may no longer justify decisions about their present. Although penal populism may make rights grounded in forgiveness politically difficult to establish, that difficulty does not remove the reasons for recognising them. As technology makes judgments easier to preserve and repeat, institutions must retain the capacity to recognise that a person can remain responsible for what they have done without being permanently defined by it.

## References

Arendt, Hannah. *The Human Condition*. Chicago: University of Chicago Press, 1958.

Donewald, Christopher, Mark Fenwick, Paulius Jurcys, Markus Lampinen, Vytautas Nekrošius, and Andrius Smaliukas. "Ownership of User-Held Data: Why Property Law Is the Right Approach." *Harvard Journal of Law & Technology Digest* (2021). *https://ssrn.com/abstract=3711017.*

Ernst, Wolfgang. *Digital Memory and the Archive*. Minneapolis: University of Minnesota Press, 2013.

Fenwick, Mark, Michael Fertik, Paulius Jurcys, and Timo Minssen. "Data Portability Revisited: Toward the Human-Centric, AI-Driven Data Ecosystems of Tomorrow." *Vanderbilt Journal of Entertainment & Technology Law* 27, no. 3 (2025): Article 7.

Fenwick, Mark, and Paulius Jurcys. "From Cyborgs to Quantified Selves: Augmenting Privacy Rights with User-Centric Technology and Design." *JIPITEC* 13 (2022): 20.

Fertik, Michael, Mark Fenwick, and Paulius Jurcys. *Data Justice: Privacy's Hidden Urgent Crisis*. Forthcoming.

Garland, David. *The Culture of Control. Oxford University Press*, 2001.

Hegel, G. W. F. *Phenomenology of Spirit*. Translated by A. V. Miller. Oxford: Oxford University Press, 1977.

Huq, Aziz Z. "Racial Equality in Algorithmic Criminal Justice." *Duke Law Journal*. 68 (2019): 1043.

Jolls, Christine, and Cass R. Sunstein. “Debiasing through Law.” *Journal of Legal Studies* 35, no. 1 (2006): 199.

Jurcys, Paulius, Chris Donewald, Jure Globocnik, and Markus Lampinen. “My Data, My Terms: A Proposal for Personal Data Use Licenses.” *Harvard Journal of Law & Technology Digest* 33 (2020).

Jurcys, Paulius, Mark Fenwick, Ashley Greenwald, Souichirou Kozuka, Valto Loikkanen, Sebastian Porsdam Mann, and Brian D. Earp. “Who Owns My AI Twin? Data Ownership in a New World of Simulated Identities.” *Computer Law & Security Review* 62 (2026): 106347.

Kirk, David S. & Sara Wakefield,‘Collateral Consequences of Punishment: A Critical Review and Pat h Forward’. *Annual Review of Criminology*, **1** (2018) 171–194.

Kleinberg, Jon, Jens Ludwig, Sendhil Mullainathan, and Cass R. Sunstein. “Discrimination in the Age of Algorithms.” *Journal of Legal Analysis* 10 (2018): 113.

Kuran, Timur, and Cass R. Sunstein. “Availability Cascades and Risk Regulation.” *Stanford Law Review* 51, no. 4 (1999): 683.

Lageson, Sarah Esther. *Digital Punishment: Privacy, Stigma, and the Harms of Data-Driven Criminal Justice*. New York: Oxford University Press, 2020.

Mayer-Schönberger, Viktor. *Delete: The Virtue of Forgetting in the Digital Age*. Princeton, NJ: Princeton University Press, 2009.

Mayson, Sandra G. “Bias in, Bias Out.” *Yale Law Journal* 128 (2019): 2218.

McNeill, Fergus, ‘Four Forms of “Offender” Rehabilitation: Towards an Interdisciplinary Perspective’. *Legal and Criminological Psychology*, 17 (2012): 18–36.

Minow, Martha. *When Should Law Forgive?* New York: W. W. Norton, 2019.

Ohm, Paul. “Broken Promises of Privacy: Responding to the Surprising Failure of Anonymization.” *UCLA Law Review* 57 (2010): 1701.

Solove, Daniel J. “A Taxonomy of Privacy.” *University of Pennsylvania Law Review* 154, no. 3 (2006): 477.

Solove, Daniel J. *Understanding Privacy*. Cambridge, MA: Harvard University Press, 2008.

Solove, Daniel J. *Privacy & Technology*. Oxford: Oxford University Press (2025a).

Solove, Daniel J. “Artificial Intelligence and Privacy.” *Florida Law Review* 77 (2025b).

Stiegler, Bernard. *Technics and Time: Volume 2*. Stanford: Stanford University Press, 2009.

Sunstein, Cass R. “Moral Heuristics.” *Behavioral and Brain Sciences* 28, no. 4 (2005): 531.

Sunstein, Cass R. “Deciding by Default.” *University of Pennsylvania Law Review* 162, no. 1 (2013): 1.

Zuboff, Shoshana. “Big Other: Surveillance Capitalism and the Prospects of an Information Civilization.” *Journal of Information Technology* 30 (2015): 75-89.